\documentclass[a4paper]{jpconf}
\usepackage{graphicx}
\usepackage{amsmath}
\usepackage{xspace}
\usepackage{cite}
\usepackage{hyperref}

\newcommand{\nuc}[2]{${}^{#2}{\rm#1}$}

\begin{document}
\title{Dilepton production at SIS energies with the GiBUU transport model}

\author{Janus Weil and Ulrich Mosel}

\address{Institut f\"ur Theoretische Physik, Universit\"at Giessen, D-35392 Giessen, Germany}

\ead{janus.weil@theo.physik.uni-giessen.de}

\begin{abstract}
 We present dilepton spectra from nucleus-nucleus collisions at SIS energies,
 which were simulated with the GiBUU transport model in a resonance-model approach.
 These spectra are compared to the data published by the HADES collaboration.
 We argue that the interpretation of dilepton spectra at SIS energies critically depends on
 the couplings between the $\rho$ meson and the baryonic resonances.
\end{abstract}


\section{Introduction}

While the vacuum properties of most hadrons are known to good
accuracy nowadays, it is a heavily debated question how these
properties change in nuclear matter. In particular, various theoretical predictions
regarding the in-medium properties of the light vector mesons have been suggested.
For recent reviews on in-medium physics, see
\cite{Leupold:2009kz,Hayano:2008vn,Rapp:2009yu}.

Among the expected in-medium effects, a so-called ``collisional
broadening'' of the meson spectral function, due to
collisions with the hadronic medium, is expected.
A second class of predictions claims that the vector-meson masses
are shifted in the medium due to the partial restoration of chiral symmetry \cite{Hatsuda:1991ez}.
QCD sum rules can constrain these effects, but do not provide definitive predictions \cite{Leupold:1997dg}.

The more prominent hadronic decay modes of the vector mesons are unfavorable
for studying in-medium effects, since they are affected by
strong final-state interactions with the hadronic medium -- in
contrast to the rare dilepton decay modes. Since the leptons only interact
electromagnetically, they are ideally suited to
carry the in-medium information outside to the detector,
nearly undisturbed by the hadronic medium.

Early measurements of dilepton spectra from heavy-ion collisions in the low-energy regime
were conducted by the DLS collaboration \cite{Porter:1997rc}, showing a clear excess over the
expected yield. A similar excess was also observed in experiments at higher energies
\cite{Adamova:2002kf,Arnaldi:2006jq}, where it could be
attributed to an in-medium broadening of spectral functions \cite{vanHees:2006ng,Ruppert:2007cr,vanHees:2007th}.
For the DLS data such in-medium effects never provided a convincing explanation - a problem
that was soon known as the ``DLS puzzle'' \cite{Bratkovskaya:1997mp,Ernst:1997yy,Bratkovskaya:1998pr,Shekhter:2003xd}.

More recently, the HADES collaboration at GSI has set up an extensive
program for measuring dilepton spectra from p+p, p+A and A+A reactions
\cite{Agakichiev:2006tg,Agakishiev:2007ts,Agakishiev:2009yf,Agakishiev:2011vf,HADES:2011ab,Agakishiev:2012tc}, in order to systematically check the old DLS data with improved statistics and to finally resolve the DLS puzzle. Up to now this endeavor has fully confirmed the validity of the DLS data and shifted the puzzle into the theory sector.


\section{The model}

In this paper, we apply the Gie\ss{}en Boltzmann-Uehling-Uhlenbeck transport model (GiBUU) \cite{Buss:2011mx,gibuu} to the nucleus-nucleus reactions studied by the HADES collaboration. The same model was used earlier to describe the elementary HADES and DLS data \cite{Weil:2012ji}. GiBUU is an hadronic transport model, which relies on a resonance-model description of $NN$ and $\pi N$ collisions at low energies.

In addition to the features described in \cite{Weil:2012ji}, the model has received a few extensions in the meantime: For the production of baron resonances via $NN\rightarrow NR$, we use angular distributions according to $d\sigma/dt\propto 1/t$, inspired by one-boson exchange, in order to improve upon the isotropic phase-space approach used earlier. At low energies, these angular distributions can become important due to the geometrical acceptance of the detector.
In the $\Delta$ Dalitz decay, we now use the form factor from \cite{Ramalho:2012ng}, which gives only moderate deviations from the assumption of a constant form factor and overall seems more reasonable than the one used previously. However, one should keep in mind that the issue of the $\Delta$ form factor has still not been settled in any way.

For the elementary dilepton spectra, of course only NN collisions played a role. In heavy-ion collisions, other types of secondary collisions can occur, such as $\pi N$ and $\pi\pi$ collisions. Details about the treatment of these processes in GiBUU can be found in \cite{Buss:2011mx}. For $\pi N$ Bremsstrahlung, we use the soft-photon approximation, in the same way as previously applied to $NN$ Bremsstrahlung \cite{Weil:2012ji}.


\section{Dilepton spectra from heavy-ion collisions}

We present here preliminary results, obtained with the GiBUU transport model, of dilepton spectra from nucleus-nucleus collisions, namely the light C+C system at beam kinetic energies of 1.0 and 2.0 AGeV and the intermediate Ar+KCl system at 1.756 AGeV, all of which have been measured by the HADES collaboration in a fixed-target setup at the SIS18 accelerator at GSI.

\begin{table}[h]
  \caption{Kinematic conditions of the collision systems measured by HADES and corresponding cuts on the single lepton momenta (all in GeV), and the maximum impact parameters in fm.}
  \begin{center}
    \begin{tabular}{|c||c|c||c|c||c|}
      \hline
      system & $E_{\rm kin}/A$ & $\sqrt{s}_{NN}$ & $p_{\rm lep}^{\rm min}$ & $p_{\rm lep}^{\rm max}$ & $b_{\rm max}$ \\
      \hline
      C + C    & 1.0   & 2.32 & 0.05 & 1.8 & 5.0 \\
      C + C    & 2.0   & 2.70 & 0.05 & 1.8 & 5.0 \\
      Ar + KCl & 1.756 & 2.61 & 0.10 & 1.1 & 6.5 \\ 
      \hline
    \end{tabular}
  \end{center}
  \label{tab:HADES_reactions}
\end{table}

Table \ref{tab:HADES_reactions} gives an overview over the different reactions, including kinematic conditions, cuts and impact parameters used. All reactions have been simulated with impact parameters restricted to the range $0\leq b\leq b_{\rm max}$
Furthermore, all spectra have been filtered through the HADES acceptance filter \cite{hades,galatyuk_priv} in order to account for the geometrical acceptance and resolution of the detector. In addition, a dilepton opening angle cut of $\theta_{ee}>9^\circ$ is applied in all cases, as well as the single-lepton momentum cuts listed in table \ref{tab:HADES_reactions}, matching the experimental analysis procedure.


\subsection{C + C}

\begin{figure}
  \begin{center}
    \includegraphics[width=\textwidth]{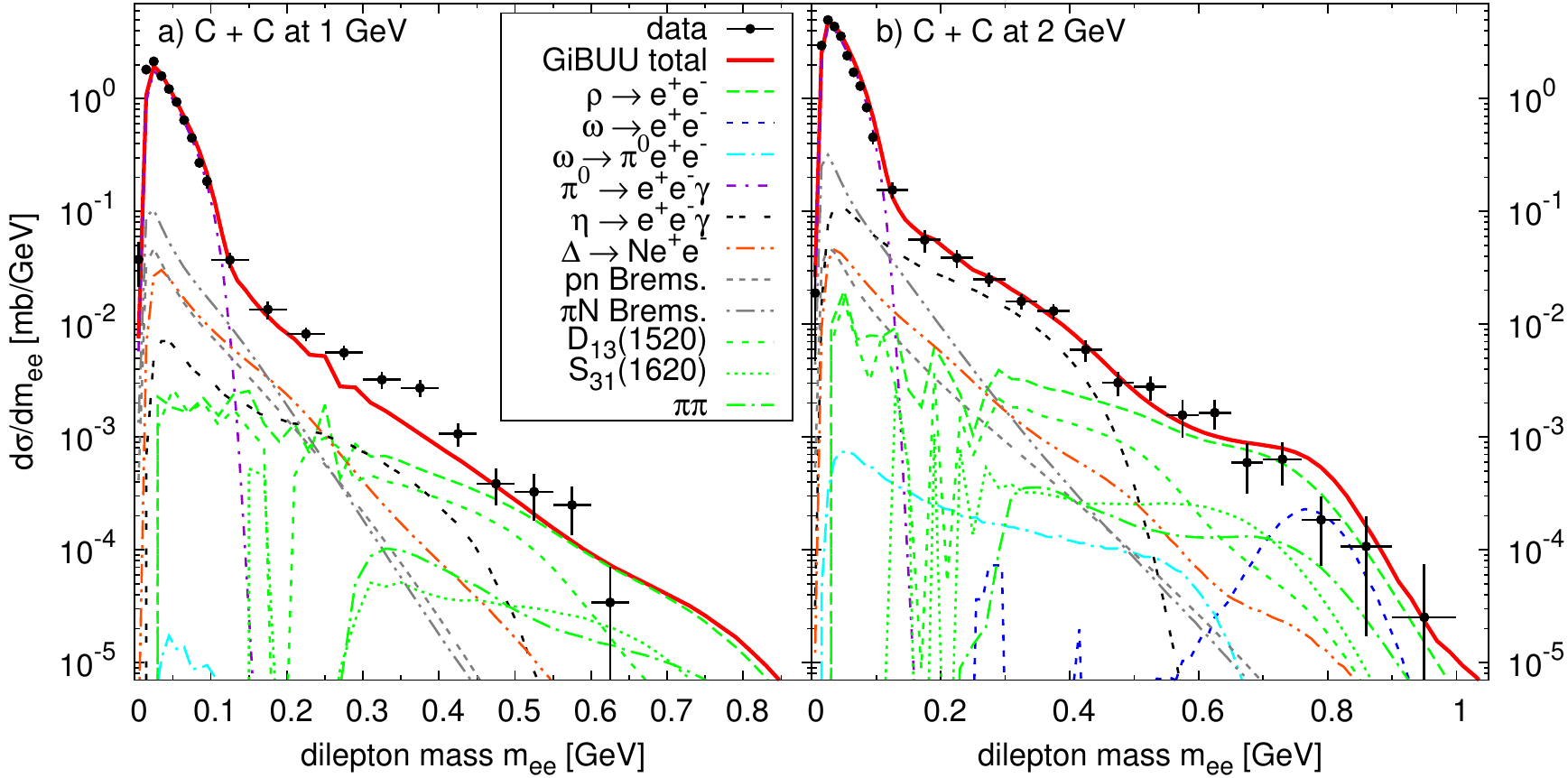}
  \end{center}
  \caption{Dilepton mass spectrum for C+C at 1 and 2 AGeV in comparison to the data from \cite{Agakichiev:2006tg,Agakishiev:2007ts}.}
  \label{fig:CC}
\end{figure}

We start by showing in fig.~\ref{fig:CC} the dilepton mass spectrum for \nuc{C}{12}+\nuc{C}{12} collisions at beam energies of 1.0 and 2.0 AGeV, compared to the HADES data (which were scaled to match the simulation in the pion channel). As has been argued in \cite{Agakishiev:2009yf}, C+C is a sufficiently light system, so that it can be regarded, in first order, as a simple superposition of elementary NN collisions, without significant secondary effects or modifications due to the hadronic medium. It has also been shown that both C+C data sets are compatible with a superposition of the NN data (when subtracting the $\eta$ component, which has a different beam energy dependence) \cite{Agakishiev:2009yf}.

From this point of view, we can expect our results for the C+C collisions to reflect the quality of our elementary NN results. And indeed we observe an excellent description of the C+C data at 2 AGeV, due to the good description of pp collisions at 2.2 and 3.5 GeV \cite{Weil:2012ji}. The largest deviation in the C+C spectrum at 2 AGeV is a slight overestimation of the data in the $\omega$ mass region, which is barely significant since the data points already have quite large error bars. Most other models showed a much more severe overestimation in this mass regime \cite{Agakichiev:2006tg,Barz:2009yz}.

In the C+C spectrum at 1 AGeV one can observe more significant deviations, namely an underestimation of the data around 300 - 400 MeV. This is probably connected to the fact that our model's agreement to the elementary data is not as good at lower energies (in particular in d+p at 1.25 GeV). Improving the description of the elementary collisions would certainly also improve the agreement with the C+C data at 1 AGeV.

We stress that, in contrast to other models \cite{Bratkovskaya:2007jk,Schmidt:2008hm}, the $\Delta$ Dalitz channel plays only a minor role in our cocktail, while the $\rho$ production via baryon resonances is much more important. Fig.~\ref{fig:CC} shows separately the dominant sources of $\rho$ mesons, which are given by decays of the $D_{13}(1520)$ and $S_{31}(1620)$ resonances and the process $\pi\pi\rightarrow\rho$. We also note that the latter is the only production process of the $\rho$ which has a physical threshold at $m=2m_\pi$. All other processes yield dilepton contributions below the $2\pi$ threshold (but unfortunately suffer from poor statistics there, due to our numerical treatment).


\subsection{Ar + KCl}

\begin{figure}
  \begin{center}
    \includegraphics[width=0.49\textwidth]{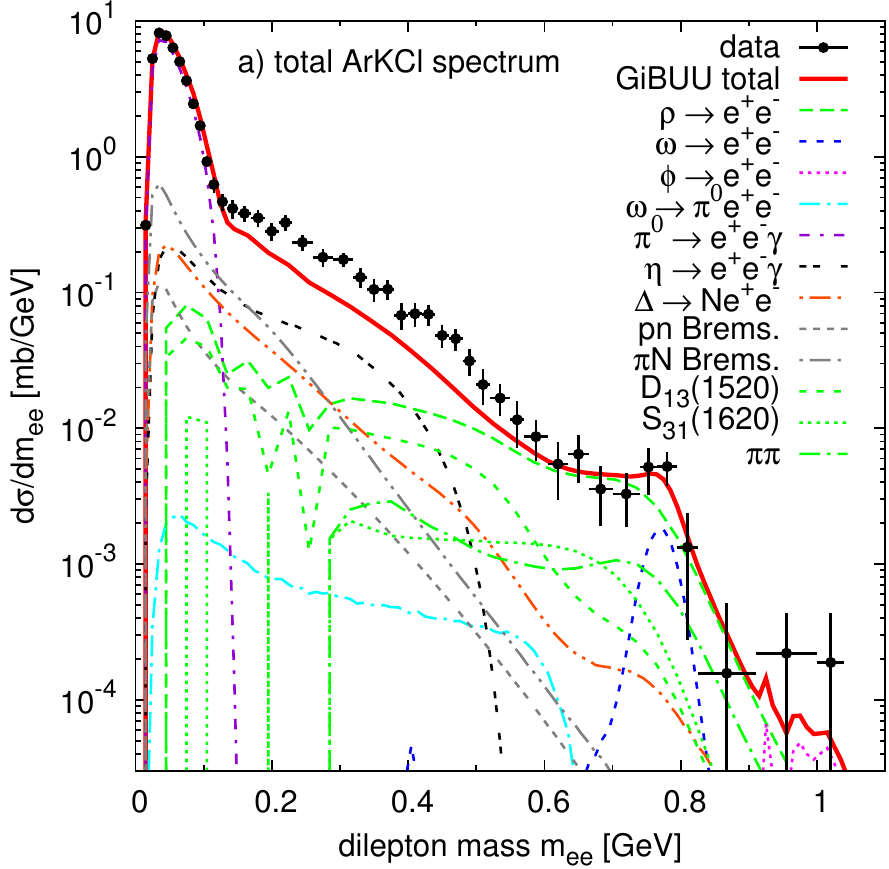}
    \includegraphics[width=0.49\textwidth]{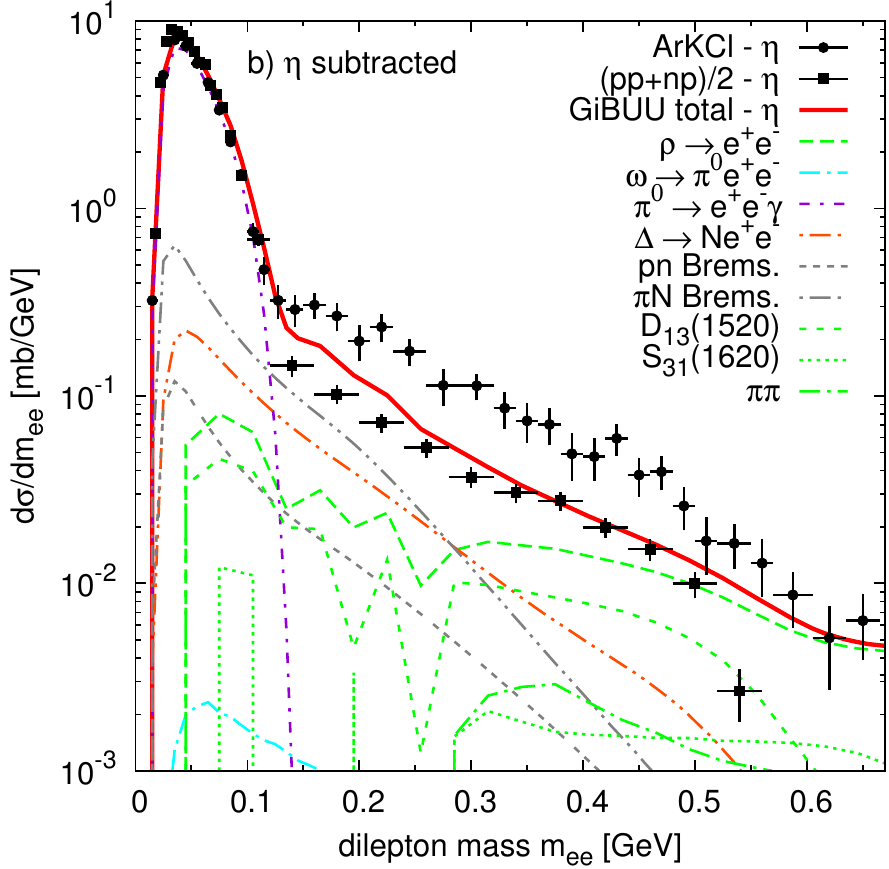}
  \end{center}
  \caption{Dilepton mass spectrum for Ar+KCl at 1.756 AGeV in comparison to the data from \cite{Agakishiev:2011vf}. Left: Total spectrum; right: $\eta$ component subtracted.}
  \label{fig:ArKCl}
\end{figure}

The data for this reaction have been obtained by shooting \nuc{Ar}{40} ions on a natural KCl target. In the simulation we use a target of \nuc{Ar}{37} instead, which represents the average of the \nuc{Cl}{35} and \nuc{K}{39} nuclei. Fig.~\ref{fig:ArKCl}a shows a comparison of our simulation results to the data, which clearly exhibits larger deviations than in the C+C case. The discrepancies mostly concern the intermediate mass range of 150 to 500 MeV.

As in the C+C case at 1 AGeV, part of the missing yield might be due to disagreement with elementary data. However, it is unlikely that this can account for all of the missing yield, since our agreement with the elementary HADES and DLS data is quite good in the energy regime relevant for the Ar+KCl system \cite{Weil:2012ji}. Instead, it seems more plausible that the missing yield represents an actual medium effect. This is supported by the fact that the ArKCl data apparently shows an excess over the elementary reference cocktail \cite{Agakishiev:2011vf}, which is of similar magnitude as the discrepancy to our model. To illustrate this, fig.~\ref{fig:ArKCl}b shows the ArKCl spectrum with the $\eta$ Dalitz component subtracted, compared to the NN reference spectrum (both normalized to the pion channel, for details see \cite{Agakishiev:2011vf}). While our model underestimates the ArKCl data, it lies close above the elementary reference spectrum.

The excess yield could in principle come from two distinct sources:
\begin{enumerate}
 \item Dilepton emission from secondary collisions (e.g. $\pi N$), which is in principle covered by our model but could be underestimated. Here also $\pi N$ Bremsstrahlung could play a role, which is treated in our model only in soft-photon approximation (SPA).
 \item An in-medium modification of the spectral functions, e.g. a broadening of the $\rho$ meson or modifications of the nucleon resonances, which are involved in the production dynamics of the $\rho$, most prominently the $D_{13}(1520)$. Such spectral modifications are not included in our current simulation, which fully relies on vacuum spectral functions.
\end{enumerate}

In order to decide if any of these two effects can contribute to the missing yield, both should be studied more closely in future investigations.


\section{Conclusions}

We have shown that the GiBUU transport model provides a reasonable description of dilepton spectra
from C+C collisions, both at 1 and 2 AGeV. At the lower beam energy, minor discrepancies show up, which might be connected to the underestimation of the d+p reaction at 1.25 GeV. At the higher beam energy of 2 AGeV, the high-mass region of the spectrum is well-described by a dominant contribution from $\rho$ mesons produced via baryonic resonances. Most other models showed a significant overestimation of the data in this region.

Regardless of the good agreement with the elementary and C+C data, our model significantly underestimates the Ar+KCl data in the intermediate mass region when using vacuum spectral functions. This excess in the data could be an indication of non-trivial effects of the hadronic medium and gives rise to the hope that even larger effects can be observed in the Au+Au system.


\ack

We thank Tetyana Galatyuk and Romain Holzmann for providing the data.
This work was supported by HIC4FAIR, HGS-HIRe and BMBF.


\section*{References}

\bibliographystyle{epj}
\bibliography{references}


\end{document}